\documentclass[aps,prb,reprint,amssymb,amsmath,superscriptaddress]{revtex4-2}
\usepackage{graphicx}
\usepackage{epstopdf}
\input{ulem.sty}

\usepackage[english]{babel}
\bibpunct{[}{]}{,}{n}{}{}
\begin{document}

\title{The manifestation of Fermi level oscillations in the magnetoresistance of HgTe quantum wells with a split spectrum}

\author{G. M.~Minkov}

\affiliation{Institute of Natural Sciences, Ural Federal University,
620002 Ekaterinburg, Russia}

\author{O.\,E.~Rut}
\affiliation{Institute of Natural Sciences, Ural Federal University,
620002 Ekaterinburg, Russia}

\author{A.\,A.~Sherstobitov}

\affiliation{Institute of Natural Sciences, Ural Federal University,
620002 Ekaterinburg, Russia}

\affiliation{M.~N.~Miheev Institute of Metal Physics of Ural Branch of
Russian Academy of Sciences, 620137 Ekaterinburg, Russia}

\author{A.\,V.~Germanenko}
\affiliation{Institute of Natural Sciences, Ural Federal University,
620002 Ekaterinburg, Russia}

\author{S.\,A.~Dvoretski}

\affiliation{Institute of Semiconductor Physics RAS, 630090
Novosibirsk, Russia}

\author{N.\,N.~Mikhailov}

\affiliation{Institute of Semiconductor Physics RAS, 630090
Novosibirsk, Russia}
\affiliation{Novosibirsk State University, Novosibirsk 630090, Russia}

\date{\today}

\begin{abstract}
Shubnikov-de Haas (SdH) oscillations and magneto-intersubband oscillations of magnetoresistance of structures with single HgTe quantum wells with a width of $(10-18)$~nm have been experimentally studied. The spectrum of the conduction band in these structures is split by the spin-orbit interaction. This leads to beats of the SdH oscillations and the appearance of low-frequency magneto-intersubband oscillations. The mutual position of the antinodes of the SdH oscillations and the maxima of the magneto-interband oscillations is unusual –- in low magnetic fields it is directly opposite to the predictions of the theory. Measurements in high magnetic fields, in which the relative amplitude of the SdH oscillations becomes greater than $0.2-0.3$, show a change in the relative position of the antinodes of the SdH oscillations and the maxima of low-frequency oscillations. Numerical calculations and additional measurements at different temperatures show that the observed effects are due to oscillations of the Fermi level in the magnetic field.
\end{abstract}


\maketitle

\section{Introduction}
\label{sec:intr}

In systems with a two-dimensional electron gas, the magnetic field normal to the plane of the two-dimensional gas causes an orbital quantization of the energy spectrum. This leads to oscillations of the density of states $\nu(B)$ at the Fermi level and, as a result, to oscillations of magnetoresistance -- Shubnikov–de Haas (SdH) oscillations. These oscillations are periodic in the reciprocal magnetic field, the frequency of oscillations ($f$) is determined by the electron density ($n$) $f=hn/(eK)$, where $K$ is the degree of degeneracy of Landau levels. When there are two branches in the energy spectrum, the electron densities in the branches ($n_1$ and $n_2$) are different and this difference is quite large, beats appear in SdH oscillations and low-frequency oscillations occur with a frequency corresponding to  $n_3=n_1-n_2$. Such oscillations, known as magneto-intersubband oscillations (MISO), occur due to transitions between branches of the energy spectrum. Such branches arise in structures with several subbands of spatial quantization, in structures with double quantum wells, in wide single quantum wells with carriers localized at each of the walls of the well due to Coulomb repulsion \cite{Polyanovs88,Leadly92,Mamani09,Mamani09-1}.

Theoretical calculations for structures with two branches of the spectrum have been made in many papers [see, for example, Ref.~\cite{Dmitriev12} and references therein]. They give the following expressions for SdH oscillations
\begin{eqnarray}
\label{eq10}
\frac{\Delta\rho^{SdH}}{\rho_D}&=&\frac{\Delta\rho^{SdH}_1}{\rho_D}+
\frac{\Delta\rho^{SdH}_2}{\rho_D} \nonumber \\
&\simeq& -\mathcal{F}\,\delta\left[
\cos{\left(\frac{2\pi f_1}{B}\right)}+\cos{\left(\frac{2\pi f_2 }{B}\right)}\right]
\nonumber \\
&=& -2\mathcal{F}\,\delta \cos{\left[\frac{\pi(f_1-f_2)}{B}\right]} \cos{\left[\frac{\pi(f_1+f_2)}{B}\right]},
\end{eqnarray}
and for MISO
\begin{equation}
\label{eq20}
\frac{\Delta\rho^\text{MISO}}{\rho_D}=\delta_1\delta_2\frac{2}{\tau_{12}}
\frac{n_1\tau_1+n_2\tau_2}{n_1+n_2}\cos{\left[\frac{2\pi(f_1-f_2)}{B}\right]},
\end{equation}
where $\rho_D=\sigma_D^{-1}$ with $\sigma_D$ as the Drude conductivity, $\delta=\exp{[-\pi/(\omega_c\tau)
]}$, $\mathcal{F}=x/\sinh(x)$, $x=2\pi^2k_BT/(\hbar\omega_c)$, $\omega_c=eB/m$, $1/\tau_1$  and $1/\tau_2$ are the scattering rates which includes
both intrasubband and intersubband scattering for the branches 1 and 2, respectively, and $1/\tau_{12}=W_{12}$ is the probability of transitions between states of different branches averaged over scattering angles.

These formulae show that the magnetic fields corresponding to the antinodes of the SdH oscillations must coincide with the fields of the MISO maxima. The amplitude of the MISO should increase monotonously with an increase in the magnetic field and be suppressed with an increase in temperature significantly less than the amplitude of the SdH oscillations. It is precisely this behavior of MISO and SdH oscillations that is observed in \cite{Polyanovs88,Leadly92,Mamani09,Mamani09-1}.

MISO behave noticeably differently in structures in which two branches of the spectrum arise due to spin-orbit interaction (SOI), which leads to the formation of two single-spin branches of the spectrum $E^{(1)}(\mathbf{k})$ and $E^{(2)}(\mathbf{k})$, where $\mathbf{k}$ is a quasimomentum, and spin-orbit splitting $\Delta_\text{SOI}=E^{(1)}(\mathbf{k})-E^{(2)}(\mathbf{k})$. This is manifested in the mutual position of the MISO maxima and the antinodes of SdH oscillations \cite{Minkov19,Minkov20-2}. It is the exact opposite of the cases \cite{Polyanovs88,Leadly92,Mamani09,Mamani09-1}.

\section{experimental}
\label{sec:exp}

Our samples with the HgTe quantum wells of different width were realized on the basis of HgTe/Hg$_{1-x}$Cd$_{x}$Te ($x=0.7$) heterostructures grown by the molecular beam epitaxy on a GaAs substrate \cite{Mikhailov06}. The samples were mesa etched into standard Hall bars of $0.5$~mm  width, the distance between the potential probes was $0.5$~mm. To change and control the carrier density in the quantum well, the field-effect transistors were fabricated with parylene as an insulator and aluminium as a gate electrode. For each heterostructure, several samples were fabricated and studied. The parameters of the structures under study are presented in the Table~\ref{tab1}. All measurements were carried out using the  {\it dc} technique in the linear response regime at $T=(1.33\ldots 10.0)$~K within the magnetic field range $(-2.0\ldots 2.0)$~T. The results obtained for all the studied structures are similar. In more detail, the results for the structure 190319 will be considered.

\begin{table}
\caption{The parameters of  heterostructures under study }
\label{tab1}
\begin{ruledtabular}
\begin{tabular}{ccccccc}
 \# & $d$ (nm) & $V_g$(V) & $n$($10^{11}$~cm$^{-2}$)& $\mu$ (cm$^2$/Vs)
&   $\Delta_\text{SOI}$ (meV) \\
\colrule
  150224& 10.0 & $4.0$ & $8.8$  & $3.6\times 10^{5}$   &
$11.6$\\
  190319& 14.1 & $5.0$ & $5.5$   & $5.0\times 10^{5}$ &    $9.5$\\
  100623& 18.0 & $3.8$ & $4.5$    & $6.5\times 10^{5}$ &
$9.9$\\
\end{tabular}
\end{ruledtabular}
\end{table}

\section{results}
\label{sec:res}

\begin{figure}[h!]\centering
\includegraphics[width=\linewidth,clip=true]{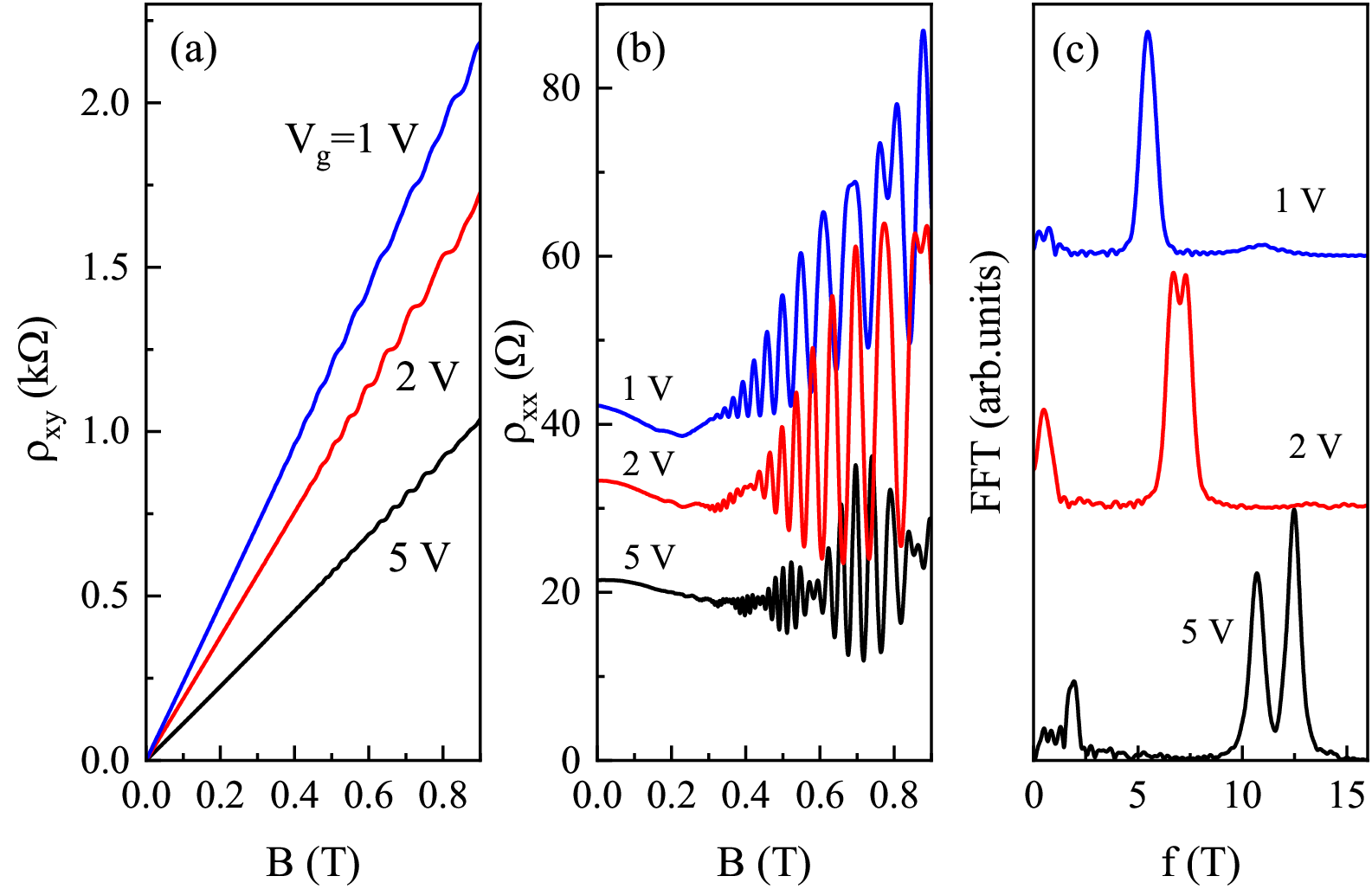}
\caption{(color online)  Magnetic field dependences of $\rho_{xx}$ (a), $\rho_{xy}$ (b), and the Fourier spectra for $\delta\rho(B)$ (c)  for several gate voltages. Structure 190319. $T=1.33$~K. }
\label{f1}
\end{figure}

The magnetic field dependences of the longitudinal ($\rho_{xx}$) and transverse ($\rho_{xy}$) magnetoresistance for some  $V_g$ values   for the structure 190319  are shown in Figs.~\ref{f1}(a) and \ref{f1}(b), respectively.  As seen from Fig.~\ref{f1}(a) $\rho_{xy}$ linearly depends on the magnetic field at $B<0.4$~T. The Hall electron density was defined as $n_H= B/(e\rho_{xy}^\text{le})$, where $\rho_{xy}^\text{le}$ is the linear interpolation of $\rho_{xy}(B)$.

\begin{figure}[h!]\centering
\includegraphics[width=\linewidth,clip=true]{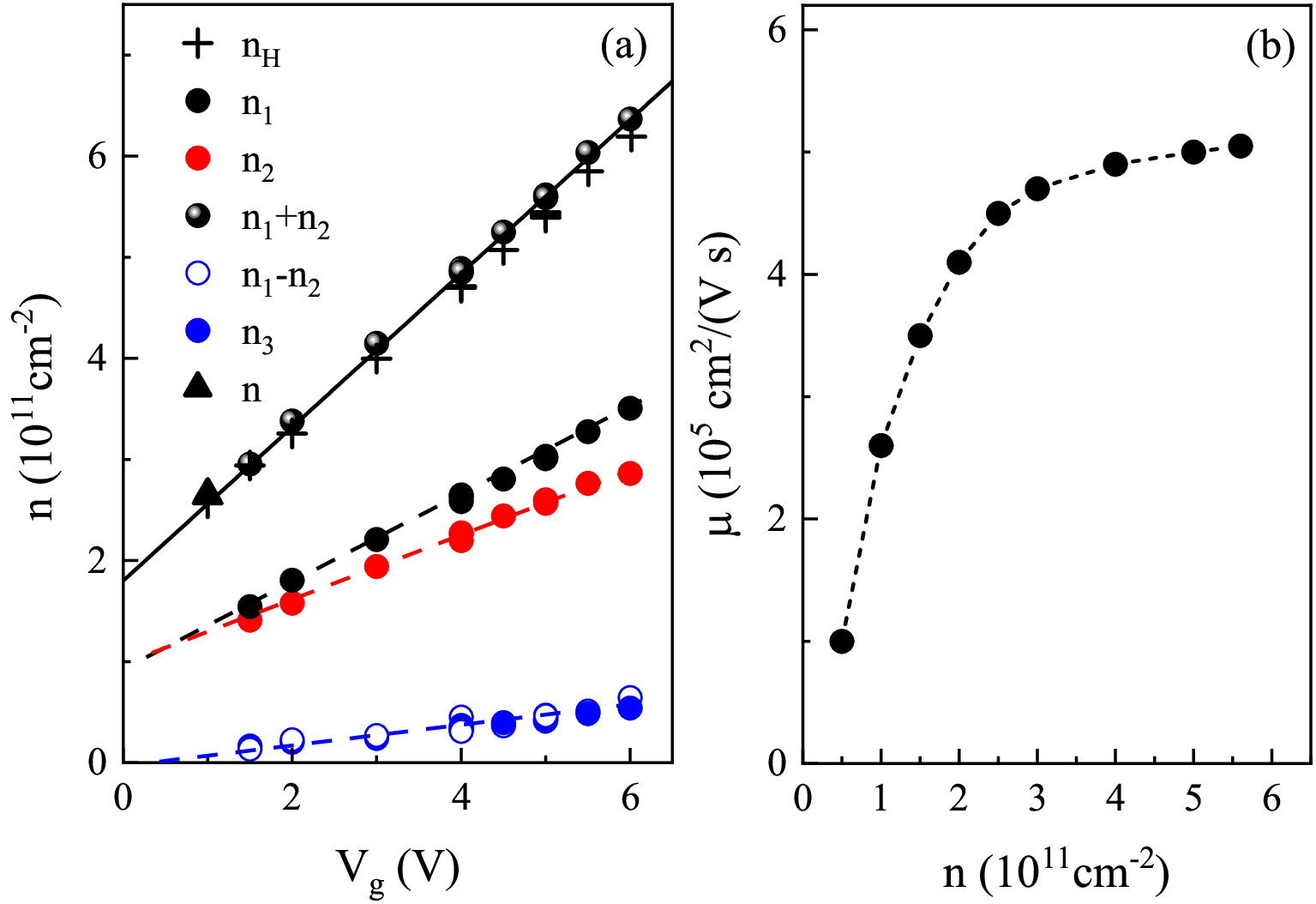}
\caption{(color online) (a) -- The gate voltage dependences of the electron density obtained from the Hall effect $n_H$ and from the Fourier spectra $n_i$, $i=1\ldots 3$. (b) -- the electron density dependence of the mobility. }
\label{f2}
\end{figure}

The Fourier analysis of the oscillating part of $\rho_{xx}(B^{-1})$ defined as $\delta\rho=(\rho_{xx}-\rho^\text{mon})/\rho^\text{mon}$, where $\rho^\text{mon}$ is the monotone part of the magnetoresistance, was performed for all the gate voltages. Fig.~\ref{f1}(c) shows that at $V_g = 1$~V, one component is observe in the Fourier spectrum, and with an increase in $V_g$ it splits into two components and, in addition, a low-frequency component appears. The interpretation of the above results becomes transparent from Fig.~\ref{f2}(a), which shows the $V_g$ dependences of the Hall electron density and the electron density determined from the Fourier spectra. At $V_g = 1$~V the electron density obtained as $n=hf/(2e)$   coincides with the Hall electron density that indicates a two-fold degeneracy of the Landau levels. For the higher $V_g$ values, when splitting of Fourier spectra is observed, the Hall electron density coincides with $n_1+n_2$, where $n_{1,2}=hf_{1,2}/e$, that indicates the degeneracy of the Landau levels is equal to one.

Figure.~\ref{f2}(a) shows that  the Hall electron density coincides with $n_1+n_2$ in the entire $V_g$ range within the experimental error. The density $n_3$ corresponding to the low-frequency component  coincides with the difference $n_1-n_2$. The monotonous increase in electron mobility with increasing electron concentration shows that only the first subband of spatial quantization is occupied in this range of $n$. The beginning of filling of the second subband is accompanied by a rather sharp decrease in the effective mobility $\mu=\rho_{xy}(B)/[B·\rho_{xx}(0)]$ at $B=0.3$~T \cite{Khudai22,Minkov22}.

Thus, the above results show that the formation of two branches of the spectrum in this structure occurs due to spin-orbit interaction. The splitting value $\Delta_\text{SOI}$ increases with an increase in the electron density due to gate voltage increase.

The beats of high-frequency oscillations (HFO) are caused by a superposition of oscillations of the magnetoresistance corresponding to two branches split due to spin–orbit interaction. The  low-frequency oscillations (LFO) are caused by electron transitions between these branches. The linearity of the dependence $n_H(V_g)$, the coincidence of the Hall electron density $n_H$ with $n_1+n_2$ found from the Fourier spectra, the proximity of $e\,dn/dV_g$ with the capacity between gate electrode and 2D electron gas measured directly, all these  show that $n$ does not depend on the magnetic field.

\begin{figure}[h!]\centering
\includegraphics[width=0.65\linewidth,clip=true]{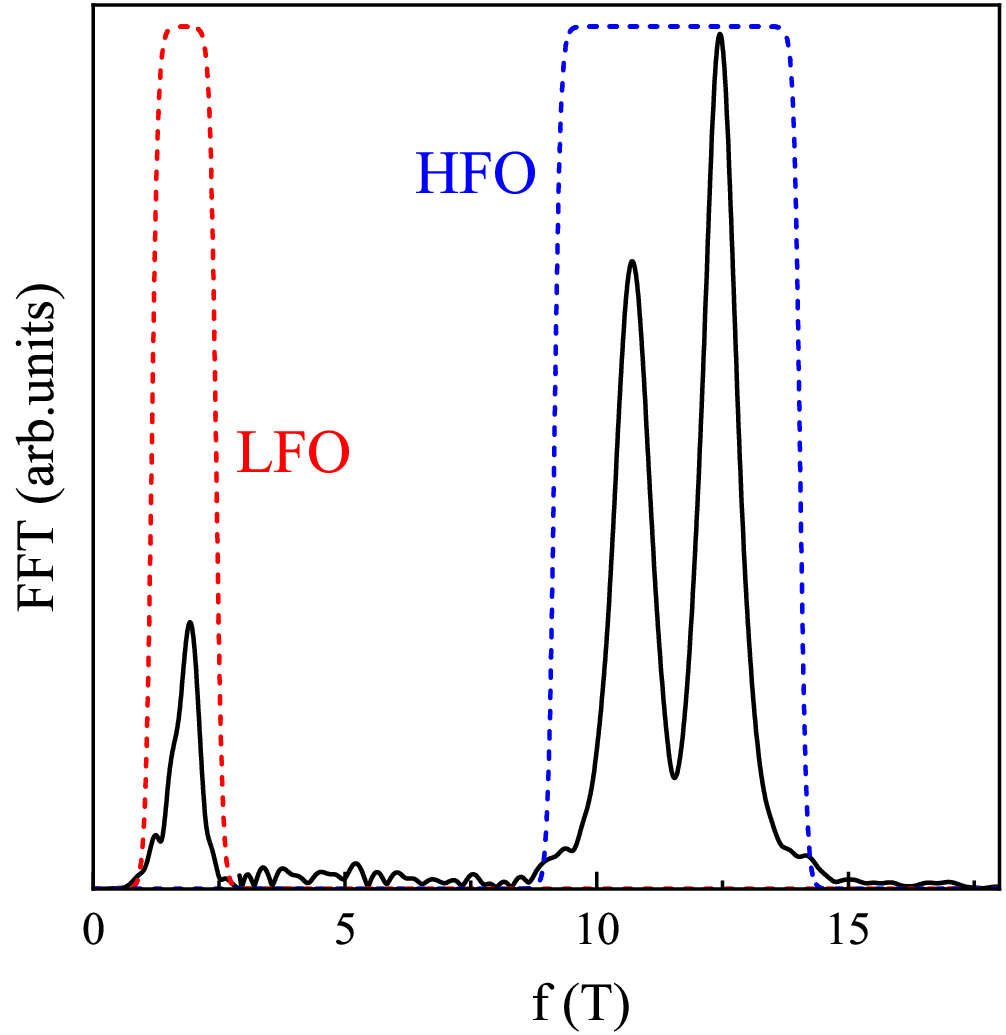}
\caption{(color online) The Fourier spectrum of $\delta\rho(B)$ for the structure 190319 at $V_g=5$~V, $n=5.5\times 10^{11}$~cm$^{-2}$ and the shape of the Fourier filters  used to separate HFO and LFO. $T=1.46$~K. }
\label{f3}
\end{figure}

\begin{figure}[h!]\centering
\includegraphics[width=\linewidth,clip=true]{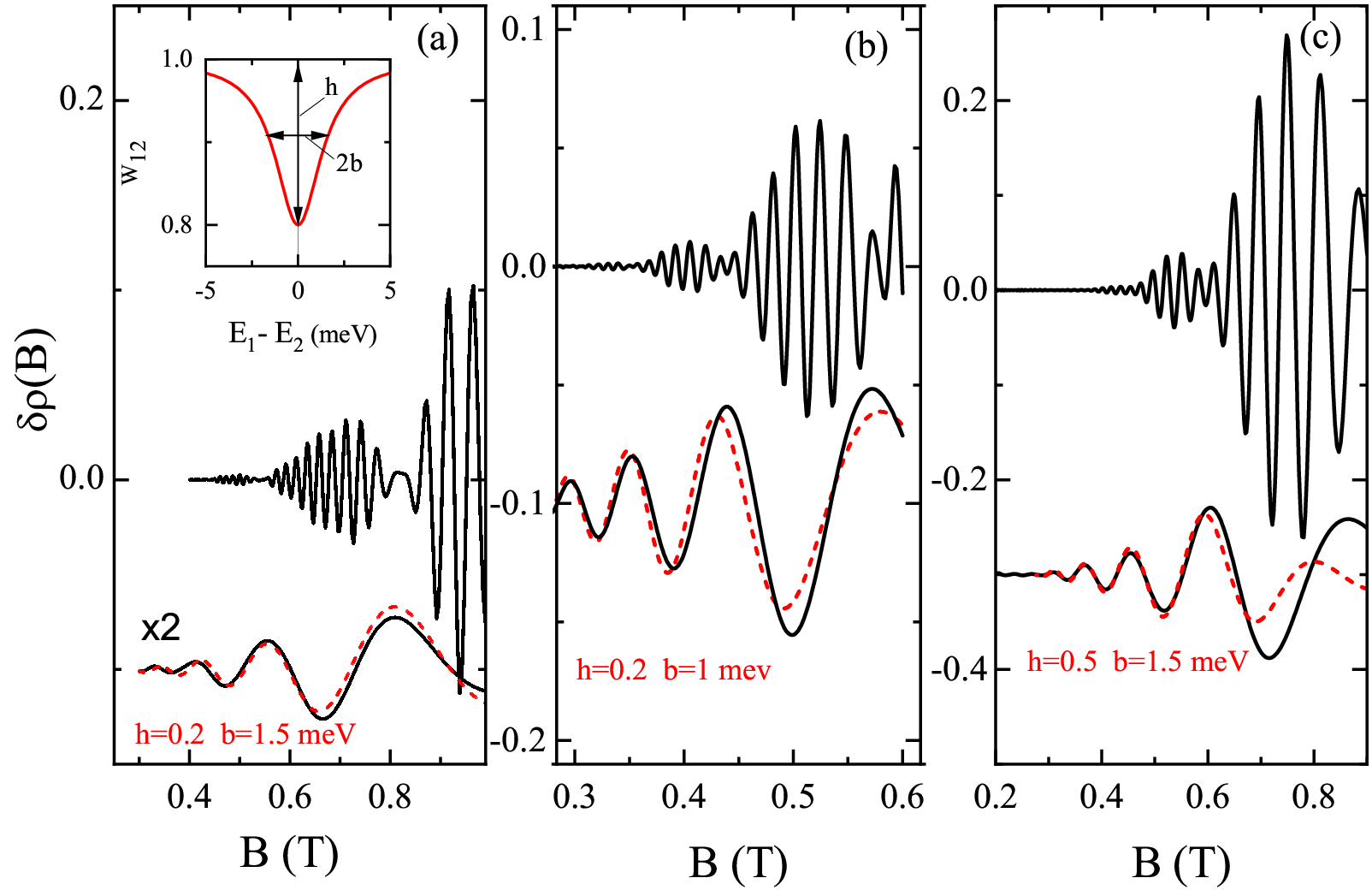}
\caption{(color online) Separated oscillations of $\delta\rho(B)$ for structures 150224 (a), 190319 (b), and 100623 (c) for the gate voltages listed in the Table~\ref{tab1}. The upper curves are HFO. The lower solid and dashed  curves are LFO  obtained by filtering of the experimental curves and  calculated in the ``toy model'' with the parameters specified in the graphs, respectively. Inset in panel (a) shows the dependence of the probability of transitions between Landau levels on the difference between the energies of the levels. $T = 4$~K. }
\label{f4}
\end{figure}

For a more detailed analysis, the Fourier filtering was used. It allows us to separate the HFO and LHO contributions. As an example, Fig.~\ref{f3} shows the Fourier spectrum of $\delta\rho(B)$ for the structure 190319 at $V_g=5.0$~V, $n=5.5\times 10^{11}$~cm$^{-2}$ and the shape of filters for separating HFO and LFO. The procedure for separating oscillations is described in detail in \cite{Minkov20-2}. After the inverse Fourier transform, we get the separated  HFO and LHO contributions  in $\delta\rho(B)$. For the three structures studied, they are shown in Fig.~\ref{f4}.

Figure ~\ref{f4} shows that  the magnetic fields of the MISO maxima coincide with the magnetic fields of the SdH oscillation nodes for all three  structures. Similar mutual positions of the MISO maxima and SdH oscillation nodes are observed in all structures studied by us with $d$ from $4.6$~nm to $46$~nm. This is exactly opposite to the theory [see expressions (\ref{eq10}) and (\ref{eq20})] and experimental results obtained for the structures with double quantum wells and with a single wide quantum well \cite{Polyanovs88,Leadly92,Mamani09,Mamani09-1}.

Note approximations were made when obtaining the expressions (\ref{eq10}) and (\ref{eq20}). They are the following:
\begin{enumerate}
\item
It was assumed that the rate of transitions between branches $1/\tau_{12}$ does not depend on either the magnetic field or the numbers of Landau levels between which transitions occur;
\item
The Fermi level oscillations with an increase in the magnetic field, which should be if the electron concentration is a constant, were neglected.
\end{enumerate}

Let us consider what changes in $\delta\rho(B)$ the approximations made can lead to. The feasibility of the first approximation  ($\tau_{12}$  is a constant) seems doubtful. In our case, transitions occur between single-spin Landau levels, so that the electron must change both the ``spin'' and the momentum (the position of the center of the oscillator). So, we can assume that $\tau_{12}$ has a minimum when the energies of the initial and final states coincide. Such a phenomenological ``toy model'' was used in \cite{Minkov20-2}. The model is based on the assumption that the probability of transition between Landau levels numbered $i$ in one subband and $j$ in another one depends on the energy difference of these levels $\Delta_{ij}=|E_i^\text{(1)}- E_j^\text{(2)}|$, and it is minimal when these energies coincide, i.e., $\Delta_{ij}=0$:
\begin{equation}\label{eq30}
  W_{ij}=\frac{1}{\tau_{12}}\left(1-h\frac{b^2}{\Delta_{ij}^2+b^2}\right).
\end{equation}
In this case, the expression for $\Delta\rho^\text{MISO}$ takes the form
\begin{eqnarray}\label{eq40}
  \frac{\Delta\rho^\text{MISO}}{\rho_D}&\sim & B^2 \sum_{i} \left\{ L(E_{i}^{(1)})\right. \nonumber \\
  &\times & \left.  \sum_{j} \left.L(E_{j}^{(2)}) W_{ij}(E_{i}^{(1)}-E_{j}^{(2)}) \right\}\right|_\text{LFC},
\end{eqnarray}
where
\begin{equation}\label{eq50}
  L(E)=\frac{\gamma}{\pi(E-E_F)^2+\gamma^2},
\end{equation}
$\gamma$ is the broadening of the Landau levels,
\begin{equation}\label{eq60}
  E_N^{(1),(2)}=\hbar\omega_c\left(N+\frac{1}{2}\right)\pm \frac{\Delta_{SOI}}{2},
\end{equation}
and $\text{LFC}$ means that $\Delta\rho^\text{MISO}/\rho_D$ is determined by the low-frequency component $f_1-f_2$ of right-hand side of Eq.~(\ref{eq40}).

\begin{figure}[h!]\centering
\includegraphics[width=0.75\linewidth,clip=true]{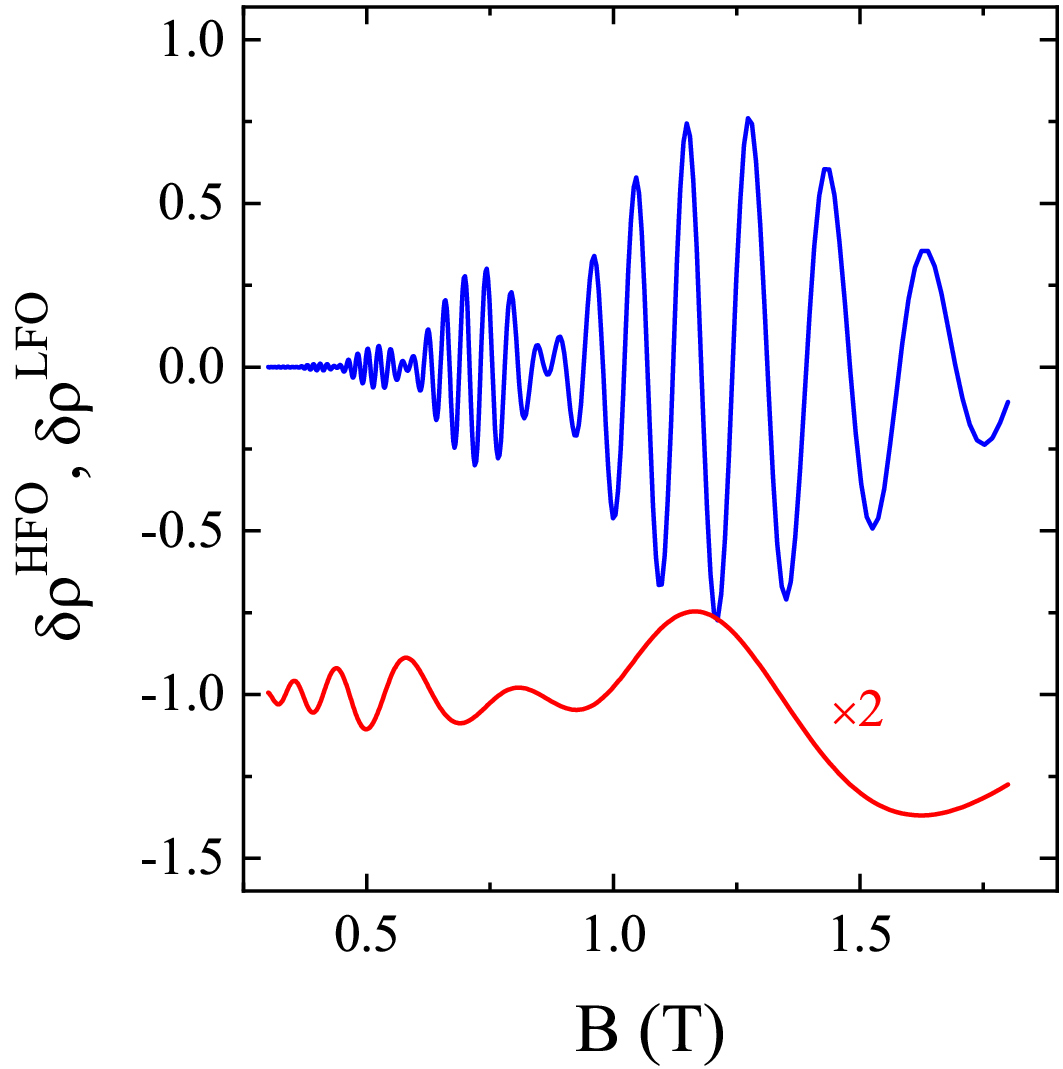}
\caption{(color online) The magnetic field dependences of the high- and low-frequency components of $\delta\rho(B)$ for the  structure 190319. $T = 4$~K. }
\label{f5}
\end{figure}

The dependences $\delta\rho^\text{MISO}(B)$ calculated within the framework of this model are shown by the dashed lines in Fig.~\ref{f4} for all  the samples under study. As seen  the mutual  positions of nodes and antinodes of SdH oscillations and maxima and minima of $\Delta\rho^\text{MISO}$ and even the dependences of $\Delta\rho^\text{MISO}(B)$ agree well with the experimental data in low magnetic fields. This good agreement is violated with an increase in $B$. It is  clearly seen in Fig.~\ref{f5} which shows the experimental dependences of $\delta\rho(B)$ for the same structure as in Fig.~\ref{f4}(b) in a wider range of magnetic fields. In the fields up to $B \simeq 0.7$~T the positions of the MISO maxima coincide with that of the nodes of the SdH oscillations, and the minima coincide with the antinodes. However, the coincidence is violated with increasing magnetic field; the MISO maximum  at $B=0.8$~T is between the antinode and the SdH oscillation node, the position of MISO maximum  at $B=1.2$~T  coincides with the position of the antinode. Note, the amplitude of the antinodes of SdH oscillations becomes relatively large in these fields (more than $0.3-0.5$). Obviously, the Fermi level oscillations can be significant in this case, that can play a role in the forming oscillation picture.

\section{Oscillations of the Fermi level}
\label{sec:Fermi}

In the case when the electron density does not change with an increase in the magnetic field, Fermi level oscillations should occur that in principle  can contribute to the dependence of the magnetoresistance on the magnetic field. Mentioning the oscillations associated with Fermi level oscillations, as a rule, it is simply said that they are negligibly small \cite{Dmitriev12}. We know  only one paper  \cite{Endo08} in which a quantitative analysis of the contribution of Fermi level oscillations to SdH oscillations  was carried out. The analysis was performed for a 2D electron gas in structures GaAs/AlGaAs with a simple spectrum. It is shown that oscillations of the Fermi level lead to only a small change in the shape of the oscillations of the density of states.

To estimate the contribution of Fermi level oscillations to the $\delta\rho$ oscillations in our case, when the electron spectrum is split, we simulated this situation. Since the carrier density (equal to the sum of the density of all states under the Fermi level) remains constant,  the $\nu$ oscillations should lead to oscillations of the Fermi level. The dependence $E_F(B)$ is given by solving equation
\begin{equation}\label{eq70}
  \int_0^{E_F(B)}\nu(E,B)dE=n,
\end{equation}
where $\nu(E,B)=\nu_0+\Delta\nu(E,B)$, $\Delta\nu(E,B)=-2\nu_0\delta\cos{[2\pi E/(\hbar \omega_c)]}$, $\delta=\exp{[-2\pi\Delta_L/(\hbar \omega_c)]}$, $\Delta_L$ is the Landau level broadening. If the spectrum is split due to SOI, the density of states is equal to the sum of the density of states in each branch
\begin{equation}\label{eq80}
  \nu(E,B)=\nu\left(E+\frac{\Delta_\text{SOI}}{2},B\right)+\nu\left(E-\frac{\Delta_\text{SOI}}{2},B\right).
\end{equation}

\begin{figure}[h!]\centering
\includegraphics[width=0.85\linewidth,clip=true]{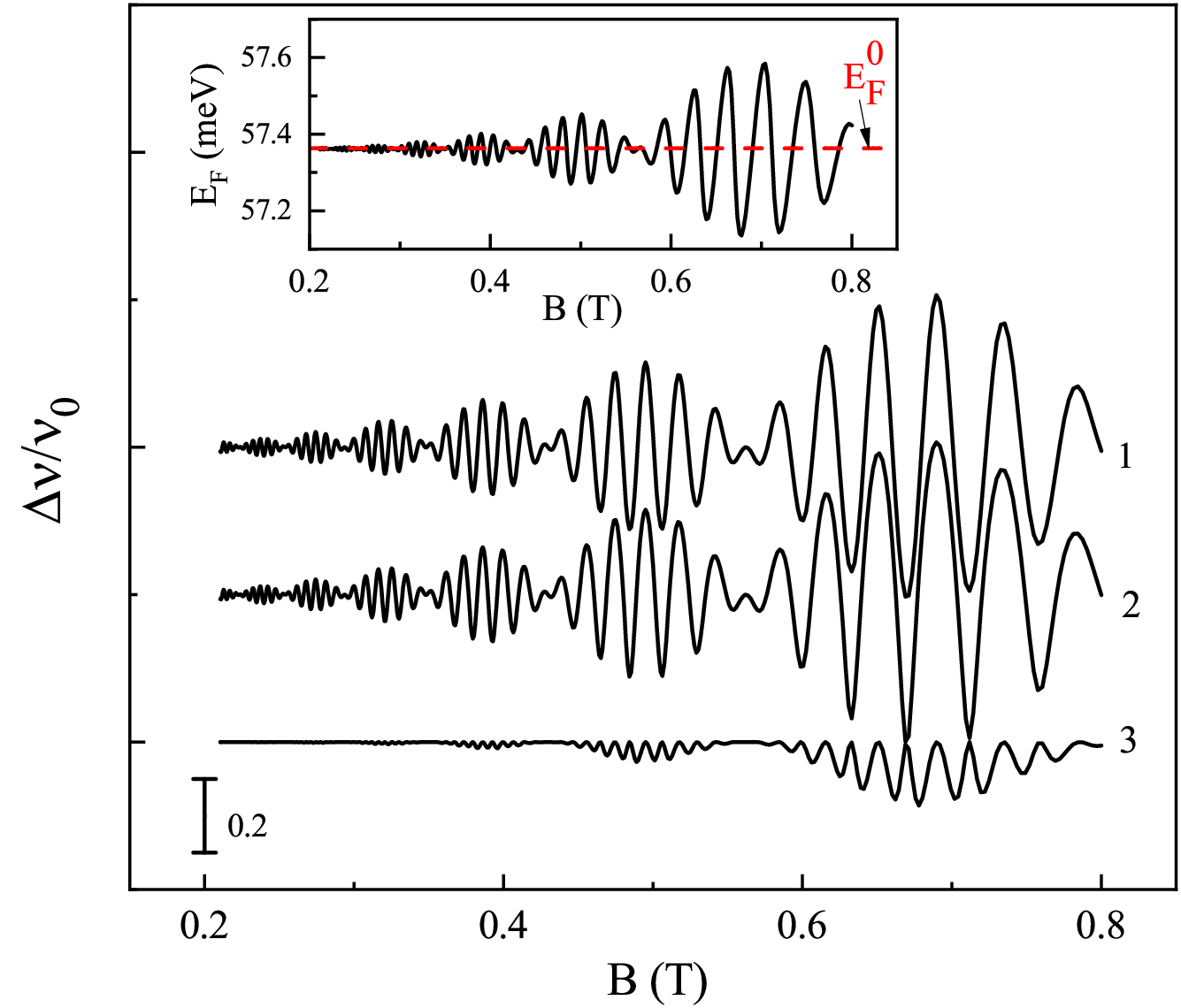}
\caption{(color online) The oscillations of the density of states calculated without (curve 1) and with (curve 2) taking into account the $E_F$ oscillations. Curve 3 is the difference between curves 2 and 1. The inset shows the $E_F$ versus $B$ dependence.}
\label{f6}
\end{figure}

Fig.~\ref{f6} shows  $\Delta\nu(B)$ calculated with parameters corresponding to the  structure 190319, for which $\delta\rho(B)$ and the Fourier spectrum are shown in Fig.~\ref{f3} and \ref{f4}(b). The inset in  Fig.~\ref{f6} shows the dependence $E_F(B)$ obtained for $n=5.5\times 10^{11}$~cm$^{-2}$ and $\Delta_\text{SOI}=9.5$~meV. Curve 1  is calculated without taking into account Fermi level oscillations, $E_F(B)=E_F^0=const$. Curve 2 is calculated taking into account Fermi level oscillations. It seems that the difference between these curves is small, nevertheless, it exists which is illustrated by curve 3.  So,  the magnetic field oscillations of the Fermi level changes the shape of the curve $\Delta\nu(B)$.

\begin{figure}[h!]\centering
\includegraphics[width=0.9\linewidth,clip=true]{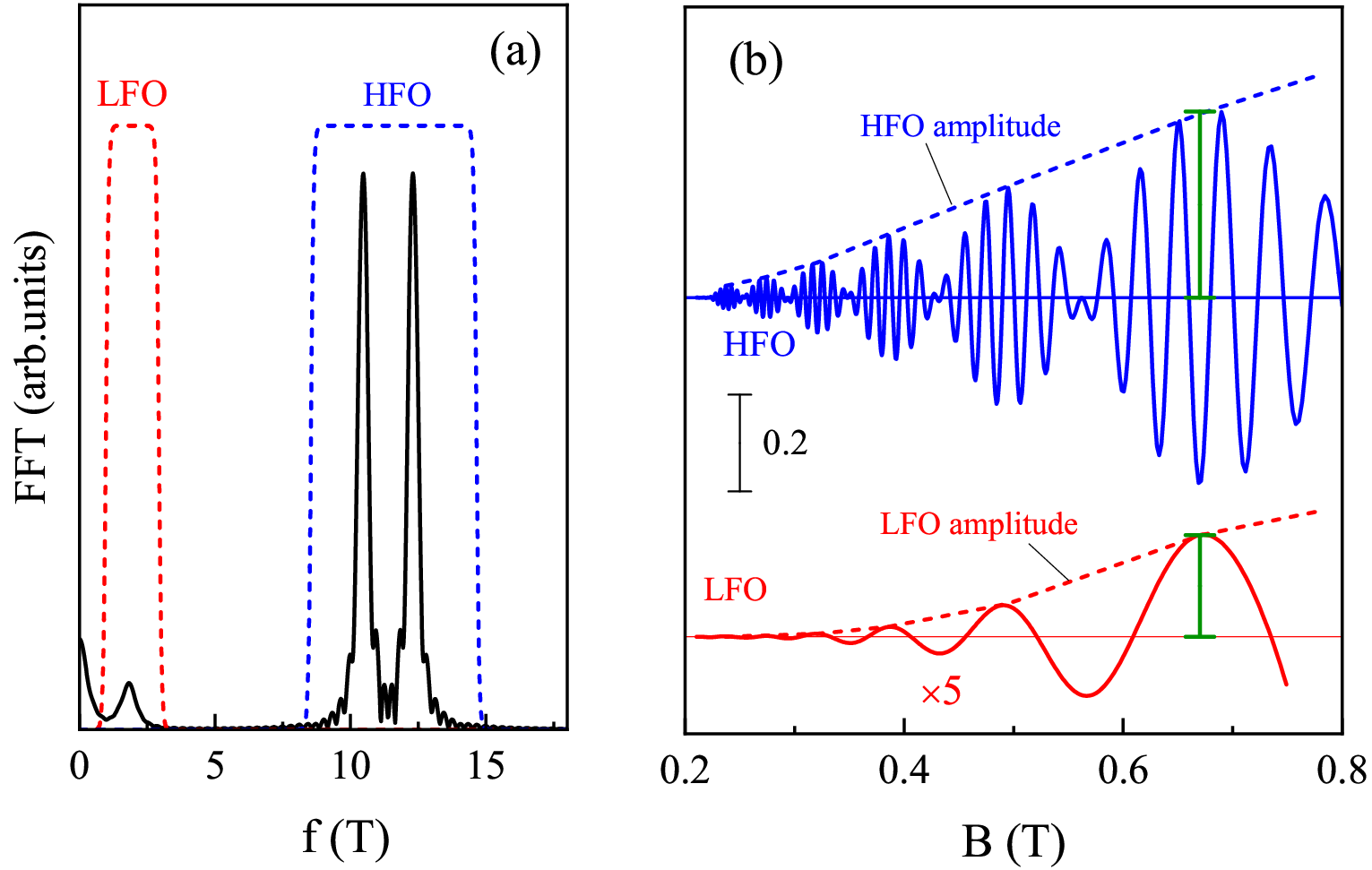}
\caption{(color online) (a) -- The result of the Fourier analysis of the magnetic field dependence of the density of states shown by curve 2 in Fig.~\ref{f6}. The dashed lines show filters for separating LFO and HFO. (b) -- HFO and LFO obtained as a result of the inverse Fourier transformation. Dashed lines and bars show the amplitude of the oscillations.}
\label{f7}
\end{figure}

Fourier analysis of curve 2 shows that Fermi level oscillations lead to the appearance of a low-frequency component having, like MISO, a difference frequency [Fig.~\ref{f7}(a)].

\begin{figure}[h!]\centering
\includegraphics[width=0.9\linewidth,clip=true]{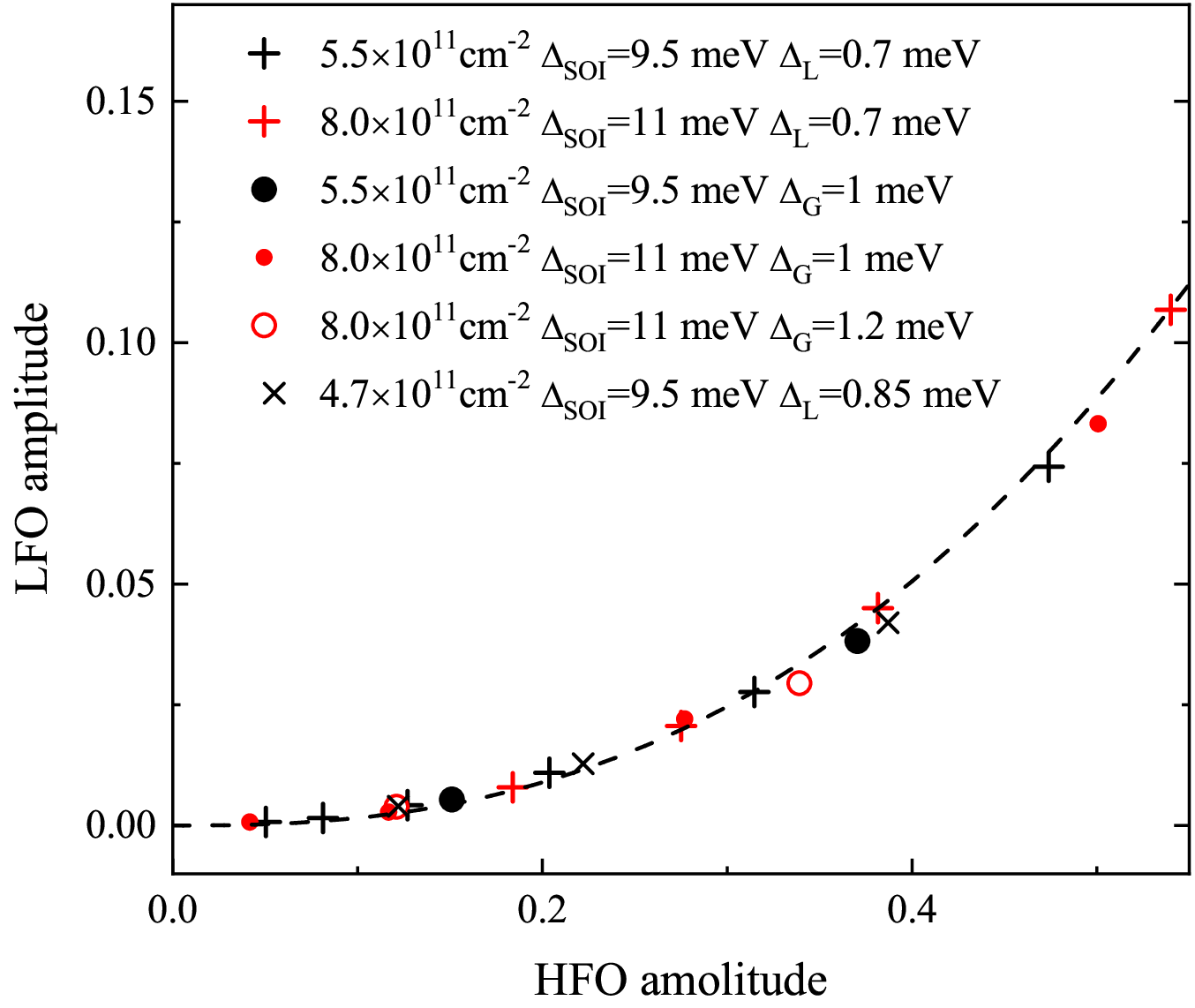}
\caption{(color online) The dependence of the LFO amplitude on the HFO amplitude of the calculated oscillation dependencies in the fields corresponding to the HFO antinodes. Different points correspond to different sets of parameters in the calculation indicated on the graph. The dashed line is a function of $y=0.5\,x^{2.5}$.}
\label{f8}
\end{figure}

The result of the Fourier analysis and the inverse Fourier transform with applying the band filters [dashed curves in Fig.~\ref{f7}(a)] are shown in Fig.~\ref{f7}(b). As firstly seen the amplitude of the LFO is significantly less than the amplitude of the HFO [see the dashed lines in Fig.~\ref{f7}(b)] \footnote{We will consider the amplitude of the oscillations to be the value corresponding to the distance from the drawn envelope to the zero line, as shown by the bars in Fig.~\ref{f7}(b).}. Secondly, the maxima of the low-frequency component are in the same fields as the antinodes in the high-frequency component. The amplitudes of the LFO maxima as a function of the  HFO amplitude in the antinode is plotted in Fig.~\ref{f8}. These dependences were obtained for different concentrations and values of spin-orbital splitting.

Similar calculations were performed for the case of Gaussian broadening of the Landau levels $\delta=\exp{[-(\pi\Delta_G/(\hbar\omega_c))^2]}$. The calculation results for different $n$, $\Delta_\text{SOI}$ and $\Delta_G$ are also shown in Fig.~\ref{f8}. As seen the LFO amplitude plotted against the HFO amplitude falls on the same curve $y(x)=0.5x^{2.5}$ as in the case of Lorentz broadening. The interpolation function allows us to estimate what the amplitude of the oscillations due to the Fermi level oscillations should be at a known HFO amplitude.

\begin{figure}[h!]\centering
\includegraphics[width=\linewidth,clip=true]{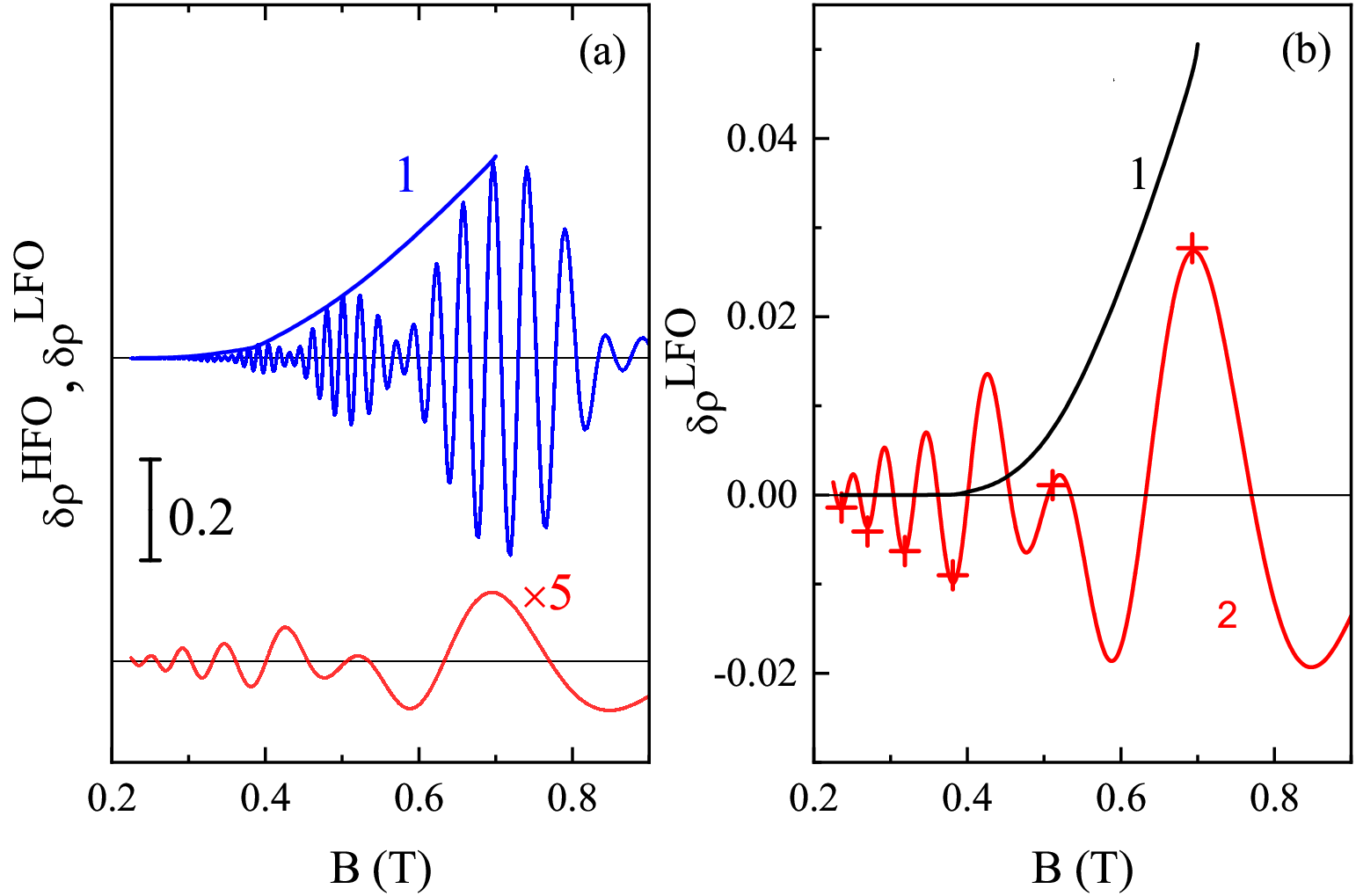}
\caption{(color online) (a) -- The oscillations $\delta\rho(B)$ obtained as a result of the inverse Fourier transform using the filters shown in Fig.~\ref{f3} for the structure 190319 at $V_g=5$~V. Curve 1 is an envelope drawn along the maxima of HFO in the antinode. (b) – Curve 1 is the amplitude of the oscillations due to the Fermi level oscillations calculated with the help of interpolation formula $y(B)=0.5[x(B)]^{2.5}$ in which the curve 1 in the panel (a) is used as $x(B)$. Curve 2 is the LFO. The crosses on curve 2 are placed in the fields corresponding to the antinodes in the HFO.}
\label{f9}
\end{figure}

Let us compare the amplitude of the oscillations associated with the Fermi level oscillations with the MISO amplitude for the structure we studied.
The interpolation formula $y(x)=0.5x^{2.5}$ allows us to construct a magnetic field dependence of the LFO amplitude due to Fermi level oscillations. The result of this procedure performed with parameters corresponding to  the structure 190319 at $V_g=5$~V is shown by the curve 1 in Fig.~\ref{f9}(b). As seen the amplitude of oscillations caused by the $E_F$ oscillations is less than $0.004$ at $B < 0.5$~T, and the oscillations are not noticeable against the background of the MISO oscillations observed in the experiment in low fields. As the magnetic field increases, the amplitude of the HFO increases and the amplitude of the oscillations caused by the $E_F$ oscillations increases sharply. When their contribution to the LFO is made more MISO, this leads to a change in the phase of the LFO.

\begin{figure}[h!]\centering
\includegraphics[width=0.75\linewidth,clip=true]{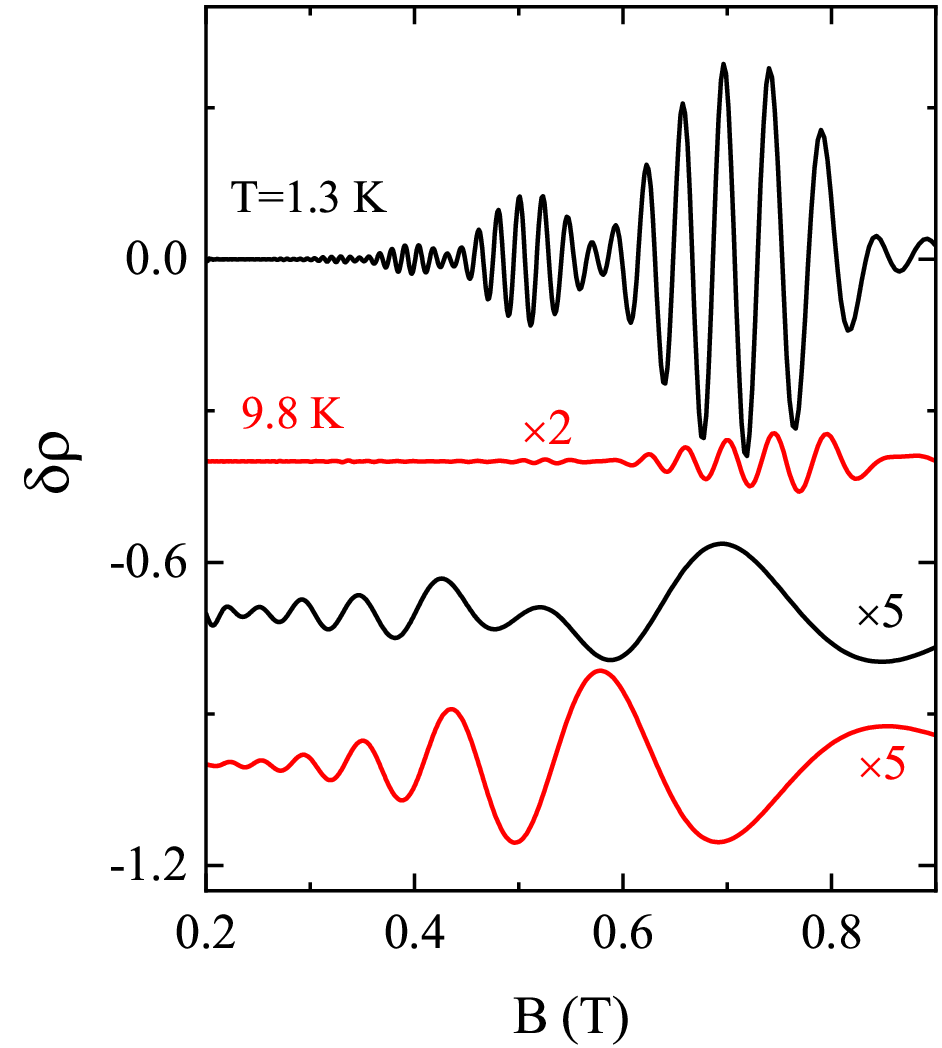}
\caption{(color online) High- and low-frequency components of $\delta\rho(B)$ for the structure 190319  at the temperatures of $1.3$~K and $9.8$~K.}
\label{f10}
\end{figure}

The validity of the above interpretation can be verified by taking measurements at low and high temperatures. Indeed, as the temperature increases, the amplitude of the HFO decreases, while the amplitude of the LFO caused by Fermi level oscillations should decrease rapidly with increasing temperature, while the MISO is weakly suppressed with increasing $T$. The results of such measurements are shown in Fig.~\ref{f10}. As seen the amplitude of the HFO decreases by an order of magnitude when the temperature increases from $1.3$~K to $9.8$~K, while the LHO amplitude changes only slightly, therewith the antinodes in the HFO correspond to the minima of the LFO. All this confirms that LFO is MISO in this case.

\section{conclusion}

Magneto-intersubband oscillations (MISO) in structures with a single HgTe quantum well with a width from $10$ to $18$~nm have been experimentally studied. Unlike structures with a double quantum well and/or a wide quantum well based on the conventional semiconductors with weak spin-orbit interaction, in HgTe quantum wells two single-spin electron branches of the energy spectrum are formed due to strong spin-orbit interaction. The presence of two branches of the energy spectrum with different carrier densities leads to beats of the SdH oscillations, which are similar in HgTe quantum wells and other quantum wells with two branches of the energy spectrum. But MISO behavior differs radically. In conventional structures, the magnetic fields of the MISO minima correspond to the antinodes of SdH oscillations, as predicted by theoretical calculations \cite{Dmitriev12}. In HgTe quantum wells, the relative position of the MISO minima relative to the antinodes of SdH oscillations is directly opposite.
We show that the unusual mutual positions of the MISO extrema and the SdH oscillation antinodes originate from the fact that the probability of transitions between the Landau levels of different branches split by SOI on the difference in their energies is dependent on the on the energy difference of the Landau levels between which the transitions occur -- the probability of transitions has a minimum when the energies of the Landau levels coincide.

The results of measurements in high magnetic fields turned out to be unexpected, when the amplitude in the antinodes of SdH oscillations becomes more than $30$~\%. Under these conditions, the position of the $\delta\rho$ maximum turns out to be in the same fields as the antinodes of the SdH oscillations. We have shown that the reason for this is the magnetic field oscillations of the Fermi level which cannot be ignored in high magnetic field.

The question may arise why the superposition of MISO and Fermi energy oscillations in structures with two branches in the spectrum have not been detected so far? The answer is simple. In conventional structures, in which two branches of the spectrum are formed due to the structure (and not due to spin-orbit interaction, as in HgTe quantum wells), the phases of MISO and oscillations due to Fermi energy oscillations are the same. Therefore, the Fermi energy oscillations change only the field dependence of low-frequency oscillations, so it is difficult to isolate this contribution in the field dependence of the $\delta\rho$ amplitude.

\begin{acknowledgments}
The work has been supported in part by the Ministry of Science and Higher Education of the Russian Federation under Project Nos FEUZ-2023-0017 and 122021000039-4 ``Electron''.
\end{acknowledgments}

%

\end{document}